\documentclass[iop,apj,revtex4]{emulateapj}

\def\lapp{\ifmmode\stackrel{<}{_{\sim}}\else$\stackrel{<}{_{\sim}}$\fi}
\def\gapp{\ifmmode\stackrel{>}{_{\sim}}\else$\stackrel{>}{_{\sim}}$\fi}

\newcommand{\tempo}{{\tt{TEMPO}}}
\newcommand{\presto}{{\tt{PRESTO}}}

\newcommand{\degrees}{^{\circ}}

\usepackage{amsmath}
\usepackage{graphicx}
\usepackage{pdflscape}
\usepackage{hyperref}

\begin{document}

\title{Timing of five millisecond pulsars discovered in the PALFA survey}

\author{
P. Scholz\altaffilmark{1},
V. M. Kaspi\altaffilmark{1},
A. G. Lyne\altaffilmark{2},
B. W. Stappers\altaffilmark{2},
S. Bogdanov\altaffilmark{3},
J. M. Cordes\altaffilmark{4},
F. Crawford\altaffilmark{5},
R. D. Ferdman\altaffilmark{1},
P. C. C. Freire\altaffilmark{6},
J. W. T. Hessels\altaffilmark{7,8},
D. R. Lorimer\altaffilmark{9},
I. H. Stairs\altaffilmark{10},
B. Allen\altaffilmark{11,12,13},
A. Brazier\altaffilmark{4,14},
F. Camilo\altaffilmark{3},
R. F. Cardoso\altaffilmark{9},
S. Chatterjee\altaffilmark{4},
J. S. Deneva\altaffilmark{15},
F. A. Jenet\altaffilmark{16},
C. Karako-Argaman\altaffilmark{1},
B. Knispel\altaffilmark{12,13},
P. Lazarus\altaffilmark{6},
K. J. Lee\altaffilmark{6},
J. van Leeuwen\altaffilmark{7,8},
R. Lynch\altaffilmark{1},
E. C. Madsen\altaffilmark{1},
M. A. McLaughlin\altaffilmark{9},
S. M. Ransom\altaffilmark{17},
X. Siemens\altaffilmark{11},
L. G. Spitler\altaffilmark{6},
K. Stovall\altaffilmark{18},
J. K. Swiggum\altaffilmark{9},
A. Venkataraman\altaffilmark{19},
and W. W. Zhu\altaffilmark{10}
}

\altaffiltext{1}{Department of Physics,
McGill University, Montreal, QC
H3A 2T8, Canada; \href{mailto:pscholz@physics.mcgill.ca}{pscholz@physics.mcgill.ca}}
\altaffiltext{2}{Jodrell Bank Centre for Astrophysics, School of Physics and Astronomy, University of Manchester, Manchester, M13 9PL, UK}
\altaffiltext{3}{Columbia Astrophysics Laboratory, Columbia University, New York, NY 10027, USA}
\altaffiltext{4}{Department of Astronomy, Cornell University, Ithaca, NY 14853, USA}
\altaffiltext{5}{Department of Physics and Astronomy, Franklin and Marshall College, Lancaster, PA 17604-3003, USA}
\altaffiltext{6}{Max-Planck-Institut f\"ur Radioastronomie, D-53121 Bonn, Germany}
\altaffiltext{7}{ASTRON, Netherlands Institute for Radio Astronomy, Postbus 2, 7990 AA, Dwingeloo, The Netherlands}
\altaffiltext{8}{Anton Pannekoek Institute for Astronomy, University of Amsterdam, Science Park 904, 1098 XH, Amsterdam, The Netherlands}
\altaffiltext{9}{Department of Physics and Astronomy, West Virginia University, Morgantown, WV 26506, USA}
\altaffiltext{10}{Department of Physics and Astronomy, University of British Columbia, 6224 Agricultural Road Vancouver, BC V6T 1Z1, Canada}
\altaffiltext{11}{Physics Department, University of Wisconsin-Milwaukee, Milwaukee WI 53211, USA}
\altaffiltext{12}{Leibniz Universit\"at Hannover, D-30167 Hannover, Germany}
\altaffiltext{13}{Max-Planck-Institut f\"ur Gravitationsphysik, D-30167 Hannover, Germany}
\altaffiltext{14}{Cornell Center for Advanced Computing, Cornell University, Ithaca, NY 14853, USA}
\altaffiltext{15}{Naval Research Laboratory, 4555 Overlook Ave SW, Washington, DC 20375, USA}
\altaffiltext{16}{Center for Gravitational Wave Astronomy, University of Texas at Brownsville, TX 78520, USA}
\altaffiltext{17}{NRAO, Charlottesville, VA 22903, USA}
\altaffiltext{18}{Department of Physics and Astronomy, University of New Mexico, NM, 87131, USA}
\altaffiltext{19}{Arecibo Observatory, HC3 Box 53995, Arecibo, PR 00612, USA}

\begin{abstract}

We present the discovery of five millisecond pulsars (MSPs) from the PALFA Galactic plane survey using Arecibo.
Four of these (PSRs~J0557+1551, J1850+0244, J1902+0300, and J1943+2210) are binary pulsars whose companions are likely white dwarfs, and
one (PSR~J1905+0453) is isolated. Phase-coherent timing solutions, ranging from $\sim$\,1 to $\sim$\,3 years in length, and based
on observations from the Jodrell Bank and Arecibo telescopes, provide precise determinations of spin, orbital, and astrometric
parameters. All five pulsars have large dispersion measures ($>100$\,pc\,cm$^{-3}$, within the top 20\% of all known Galactic field MSPs)
 and are faint (1.4\,GHz flux density \lapp\,0.1\,mJy, within the faintest 5\% of all known Galactic field MSPs),
illustrating PALFA's ability to find increasingly faint, distant MSPs in the Galactic plane. 
In particular, PSR~J1850+0244 has a dispersion measure of 540\,pc\,cm$^{-3}$, the highest of all known MSPs.
Such distant, faint MSPs are important input for accurately modeling the total Galactic MSP population.

\end{abstract}


\section{Introduction}

Millisecond pulsars (MSPs) are neutron stars that have been spun up by accretion to
extremely fast spin periods ($P<20$\,ms) \citep{acrs82,rs82}. The first discovered MSP, PSR~B1937+21, was found
in 1982 \citep{bkh+82} and since then $\sim$160 Galactic MSPs 
 have been discovered ($\sim$130 MSPs have been found in globular clusters) in different
pulsar surveys \citep[for a review of past and ongoing searches for MSPs see][]{sll13}. 
MSPs are of great scientific value, giving us insight into many different branches of fundamental
physics and astrophysics, as we elaborate below. 

Most MSPs are found in binary systems with a low-mass, degenerate companion.  These companions are 
responsible for spinning-up such pulsars to millisecond rotational periods by transferring mass and angular momentum onto them.
The discovery of new MSP binaries can test (and in some cases confirm) models of binary evolution
\citep[e.g.][]{pk94,ts99,asr+09,pfb+13} and sometimes challenge them, as in the case of PSR~J1903+0327 \citep{crl+08},
which suggested that formation in a triple stellar system occurs in some cases \citep{fbw+11,pvvn11}. 
This was also directly demonstrated in the recently discovered PSR J0337+1715 triple system \citep{rsa+14}.
On the other hand, isolated MSPs exist as well, and their origin is an important open question \citep{bg90,le91,bbb+11}.

The continued discovery of MSPs is also motivated by their potential for aiding the detection of
gravitational waves. 
High-precision, long-term timing of MSPs forming a `pulsar timing array' \citep{fb90}
could allow the detection of gravitational waves in the nanohertz frequency range particularily
in the form of a stochastic background that is due to super-massive black-hole mergers at high
redshifts \citep{vlj+11,ych+11,dfg+13,src+14}.
The discovery of more MSPs that can be timed with very high precision is crucial for this effort to succeed.
Some MSP binaries also enable tests of General Relativity and alternate theories of relativistic
gravity \citep[e.g.][]{gsf+11,fwe+12,afw+13} as well as measure precise component masses that 
can test models of neutron-star interiors \citep[e.g.][]{dpr+10,lp10,bhz+12}.

PALFA is a pulsar survey using the 7-beam Arecibo L-Band Feed Array (ALFA) on
the Arecibo Observatory William E. Gordon 305-m Telescope. The survey's centre frequency is
1.4 GHz, and the planned survey
coverage is the region of the Galactic plane ($|b| < 5\degrees$) accessible from the
Arecibo telescope ($32\degrees\lapp l \lapp 77\degrees$ and $168\degrees\lapp l \lapp 214\degrees$).
The survey began in 2004 and is ongoing \citep{cfl+06,laz13}.

\begin{figure}
\plotone{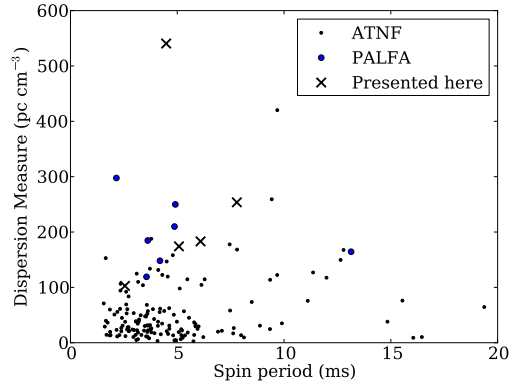}
\figcaption{ Period vs dispersion measure for all known Galactic MSPs (cf.\ Figure 5 of \citealp{csl+12}).
PALFA provides a unique window at high DMs and fast spin periods.
The PALFA pulsars are from \citet{crl+08,dfc+12,csl+12}.
The `ATNF' MSPs are for all Galactic MSPs listed in the ATNF pulsar catalog \citep{mhth05}\protect\footnotemark. 
\label{fig:pdm}
}
\end{figure}

PALFA provides a unique window for the discovery of MSPs. Its high sensitivity allows observations with
short integration times which mitigate the effects of Doppler smearing of the pulsed signal in the case of a
tight binary orbit \citep{jk91}.
The increased sensitivity and relatively high observing frequency -- which reduces
the deleterious effects of dispersion-smearing and multi-path scattering -- compared
to other surveys also allow PALFA to find MSPs at much further distances and higher dispersion measures (DMs).
Figure \ref{fig:pdm} shows the unique parameter space probed by PALFA. 

The discovery of distant, highly dispersed MSPs will allow a more complete characterization
of the Galactic MSP population. Current models of the Galactic MSP population are based primarily on the Parkes
Multibeam Pulsar Survey \citep[PMPS;][]{lfl+06} which is significantly less sensitive than PALFA and therefore biased towards nearby sources. 
The increased sensitivity of PALFA not only allows higher DMs, and thus
distances, to be probed, but allows discoveries of nearby pulsars to a lower luminosity. 
The MSPs discovered in PALFA will therefore provide valuable input to models of the Galactic MSP population
- especially the population that is confined to the Galactic plane.

High-DM MSPs also provide unique probes into the properties of the interstellar medium (ISM). As pulsar signals propagate through
the ISM, their signals are both dispersed (frequency-dependant delay), and scattered (multi-path propogation). 
These effects are time variable because of the pulsar's proper motion and thus our changing line-of-sight through the ISM.
Detailed dispersion and scattering measurements of signals from high-DM MSPs can be used to compare the observed effects to the
predicted, thus improving our models of scattering and time-dependant DM variations. 
This is important for the detection of gravitional waves, as
correcting the effects of DM variations and scattering can greatly increase the precision with which we time some MSPs \citep{kcs+13}.

PALFA has discovered 18 Galactic plane MSPs to date, including seven that have been published in previous works 
\citep{crl+08,dfc+12,csl+12}. These include: 
PSR~J1950+2414, an eccentric MSP binary that may have formed via accretion induced collapse \citep[Knipsel et al.\ in prep;][]{ft14}; 
and two MSPs for which Shapiro delay measurements were made, PSR~J1903+0327 \citep{crl+08} and PSR~J1949+3106 \citep{dfc+12}.

In this paper we present five PALFA MSP discoveries:
PSRs~J0557+1551, J1850+0244, J1902+0300, J1905+0453, and J1943+2210.
The remaining six unpublished PALFA-discovered MSPs  
will be presented in forthcoming papers along with future discoveries.
Section \ref{sec:obs} describes the discoveries as well as the follow-up timing observations and analysis.
In Section \ref{sec:disc} we discuss the results of our timing analysis and explore
the interesting properties of these newly discovered pulsars.
In Section 4, we summarize our conclusions.

\footnotetext{\url{http://www.atnf.csiro.au/research/pulsar/psrcat/}} 

\section{Observations \& Analysis}
\label{sec:obs}

\subsection{Discovery}
\label{sec:disc_obs}

The five MSPs presented here were discovered in PALFA survey observations taken between 2011 September and 2013 June.
The PALFA survey currently uses the Mock spectrometers\footnote{\url{http://www.naic.edu/~astro/mock.shtml}} as a backend for the
7-beam ALFA receiver at a central frequency of 1375\,MHz. 
The Mock backend provides 322.617 MHz of bandwidth and a time resolution of 65.476\,$\mu$s for each beam of ALFA.
PALFA has survey regions towards both the inner and outer Galactic plane.
The outer-galaxy pointings are provided by our commensal partners, the ALFA Zone-of-Avoidance Survey, 
who are mapping extragalactic neutral hydrogen emission behind the Galactic plane \citep{hsm+10}.
A typical inner-galaxy survey pointing is 268\,s and the outer-galaxy pointings
are typically 180\,s in length.
Four of the five pulsars presented here were discovered
using the above setup with the Mock backend in the inner galaxy.
PSR~J0557+1551 was discovered with the Mock backend in an outer-galaxy pointing.
The discovery observation dates for each pulsar are listed in Table \ref{ta:params}.

PALFA survey observations are processed by several pipelines. One of these is based
on the \presto\ software package \citep{rem02}\footnote{\url{http://www.cv.nrao.edu/~sransom/presto/}} 
and is described in Lazarus et al. (in prep). %\citet{patricksoon}.
In brief, the \presto-based pipeline operates on raw data automatically downloaded from a PALFA-administered archive 
at the Cornell Center for Advanced Computing and is processed on Guillimin, a Compute Canada/Calcul Qu\'ebec facility
operated by McGill University. Some processing was also performed on dedicated computer clusters at the University of British Columbia and McGill University. 
Searches are performed both in the Fourier domain, for periodic sources, and in the time domain, for single pulses, using
standard \presto\ tools.
PALFA observations are also searched by the Einstein@Home pulsar search pipeline which uses distributed volunteer computing and is described in \citet{akc+13}.
Lastly, a `quicklook' pipeline (also \presto-based), which searches reduced-resolution data in near realtime at the telescope, is also employed. 
Four of the pulsars in this work were discovered using the full-resolution \presto-based pipeline
and one, PSR~J1943+2210, was discovered with the quicklook pipeline.

\subsection{Timing Observations}
\label{sec:time_obs}

Once confirmed as pulsars, we commenced timing observations of the five new MSPs in order
to determine their rotational ephemerides.
Timing observations began at the Lovell telescope at Jodrell Bank Observatory shortly after 
discovery and Arecibo observations began up to a year later.

Follow-up observations at Jodrell Bank used a dual-polarization
cryogenic receiver on the 76-m Lovell telescope, having a system
equivalent flux density of 25\,Jy.  Data were processed by a
digital filterbank which covered the frequency band between 1350\,MHz and
1700\,MHz with channels of 0.5-MHz bandwidth. Observations were typically
made with a total duration of 10 -- 60 minutes, depending
upon the discovery signal-to-noise ratio. Data were folded at the nominal
topocentric period of the pulsar for subintegration times of 10 seconds.
After ``cleaning'' of any radio-frequency interference (RFI) using a median zapping algorithm, 
followed by visual inspection and manual removal of any remaining RFI,
the profiles were incoherently dedispersed at the nominal value of the pulsar DM.  Initial
pulsar parameters were established by conducting local searches in period
and DM about the nominal discovery values and finally summed over
frequency and time to produce integrated profiles.

The Arecibo observations were
performed with the L-wide receiver which has a frequency range of 1150--1730\,MHz and a system equivalent flux density of 3\,Jy. 
The Puerto Rico Ultimate Pulsar Processsing Instrument (PUPPI)
backend was used to record the data. Initially, each pulsar was observed in incoherent search mode with PUPPI, where
a full 800-MHz-wide spectrum in 2048 channels is read out every 40.96 $\mu$s. This produces a high
data volume, so each observation was post-processed by downsampling to 128 channels while correcting for delays due to 
dispersive effects from the interstellar medium
within each channel before summing. Later observations were performed
with PUPPI in coherent-dedispersion mode (hereafter referred to as coherent mode) 
where pulse profiles were folded in real-time and recorded in 10-s integrations
with 512 frequency channels. 
Data from outside of the 1150--1750\,MHz L-wide band were removed.
Radio frequency interference was excised with both automatic algorithms and 
manually using the {\tt PSRCHIVE} \citep{hvm04}\footnote{\url{http://psrchive.sourceforge.net/}} tools 
{\tt paz} and {\tt psrzap}.
Typically, each pulsar was observed for 10 min per observing session, 
except for PSR~J0557+1551 which was observed
for $\sim$30~min--1~hour per session.

\subsection{Timing Analysis}
\label{sec:timing}

\begin{figure*}
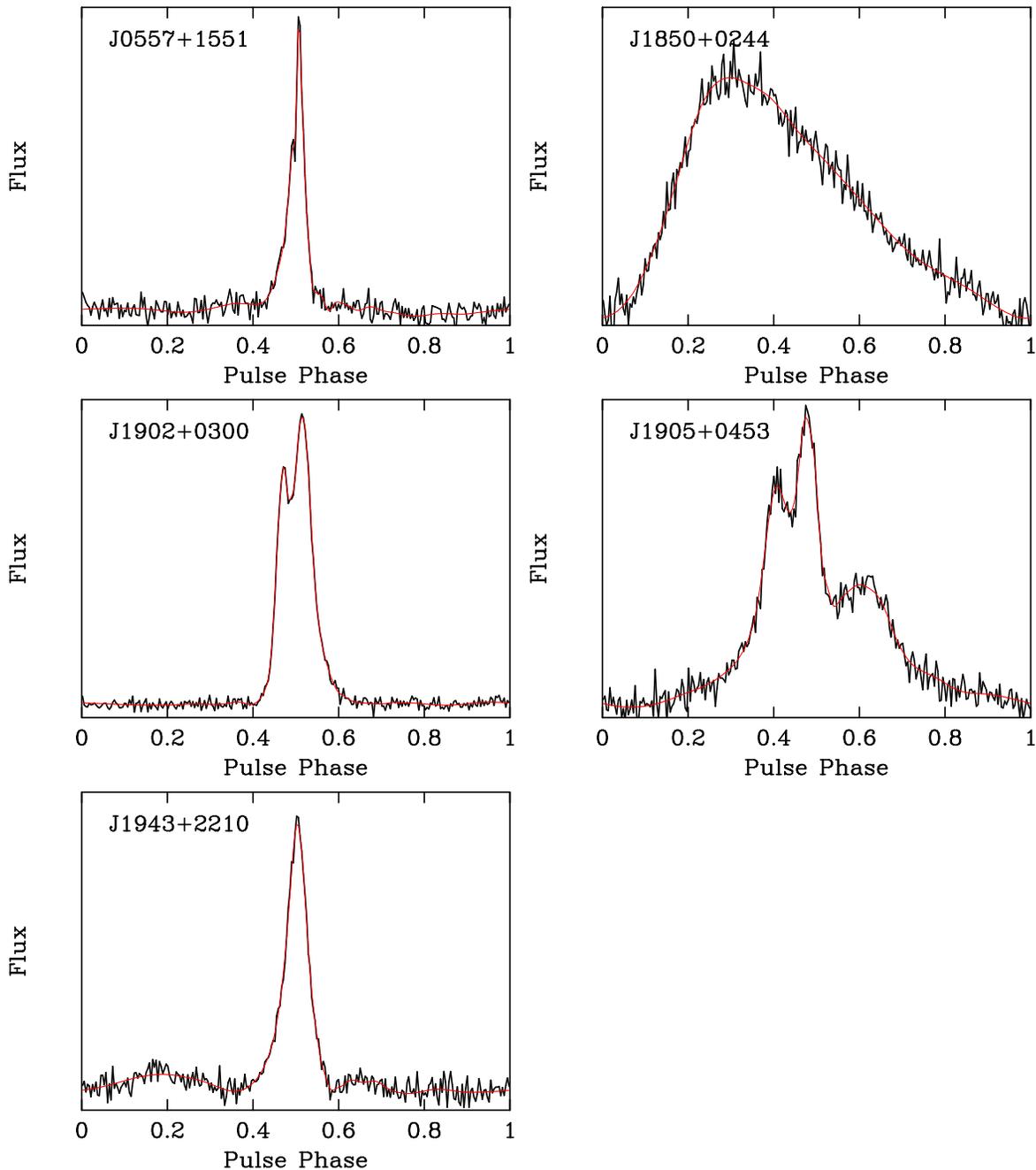

$\begin{array}{cc}
\includegraphics[angle=-90,width=3in]{0557+1551_256bin.ps} & \includegraphics[angle=-90,width=3in]{1850+0244.ps} \\
\includegraphics[angle=-90,width=3in]{1902+0300.ps} & \includegraphics[angle=-90,width=3in]{1905+04.ps} \\
\includegraphics[angle=-90,width=3in]{1943+2210.ps} \\
\end{array}$
\figcaption{ 1.4-GHz pulse profiles for the MSPs presented here. One full period is shown in 256 bins for each pulsar. 
The profiles were constructed by phase aligning and summing 
the Arecibo observations. The heavy black line shows the summed profiles and the light red line shows the profiles after smoothing was
applied. The smoothed profiles were used as template for the generation of TOAs.
\label{fig:profs}
}
\end{figure*}

\begin{figure*}
\plotone{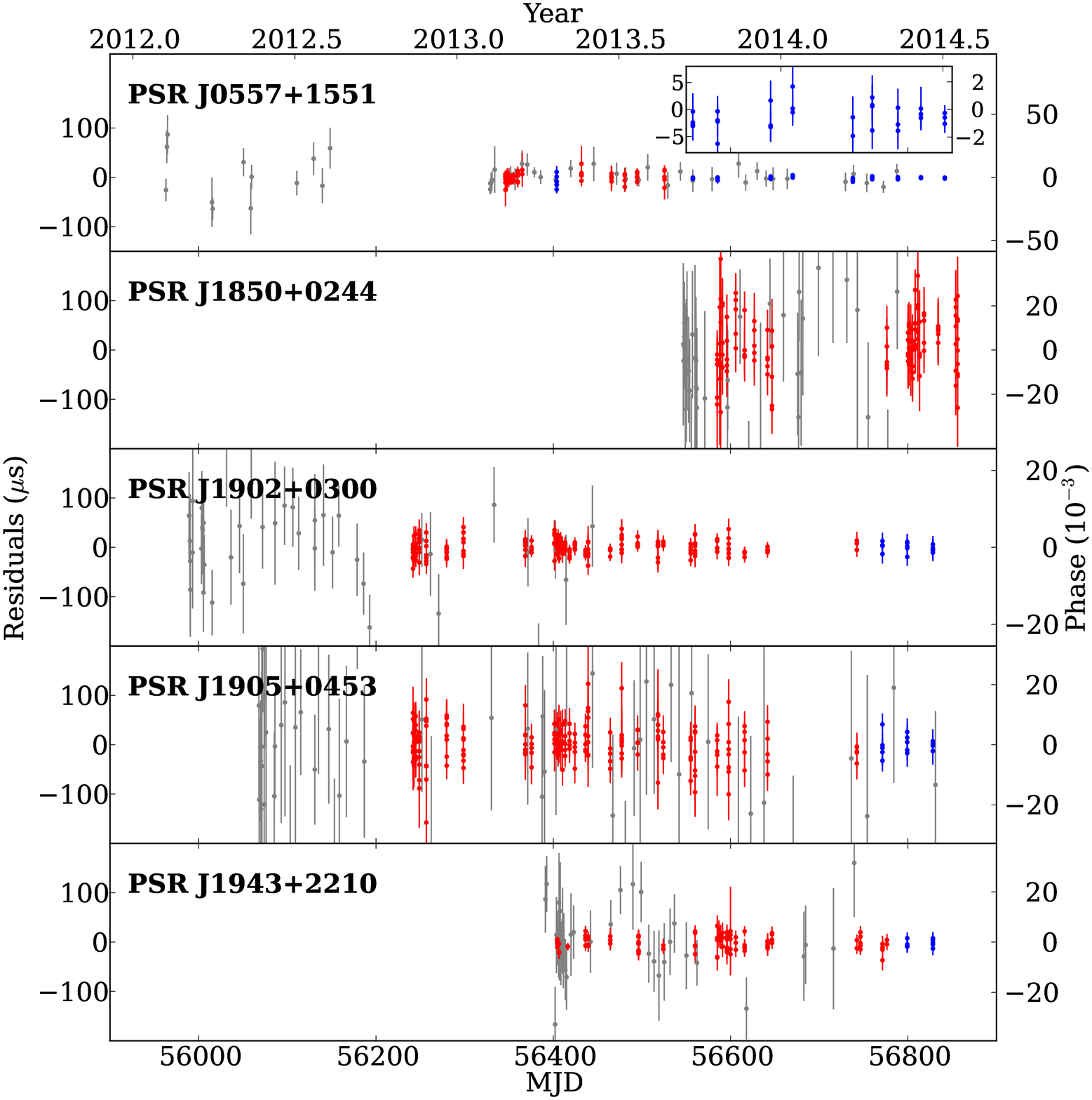}
\figcaption{ Timing residuals for the five MSPs. Gray points indicate Jodrell Bank TOAs, red points denote PUPPI search-mode TOAs
and blue points represent PUPPI coherent-mode TOAs. 
The inset shows a zoom-in of the Arecibo coherent-mode observations for PSR~J0557+1551. The x-axes of the inset are aligned with those 
of the main plot.
Error bars are scaled so that the best-fit reduced $\chi^2$ is equal to unity. The scaling factors
along with rms values for the residuals are presented in Table \ref{ta:params}.
\label{fig:resids}
}
\end{figure*}

Initial timing solutions were obtained using solely Jodrell Bank observations.
For the binary MSPs, orbital solutions were found
by measuring the frequency of the pulsar in each observation and fitting
a model of orbital Doppler shifts for a circular orbit. 
These orbital solutions were used as a starting point in the iterative timing procedure.
Pulse times-of-arrival (TOAs) were obtained after matching with a standard template constructed from the Jodell Bank
observations and processed using standard pulsar timing techniques 
with {\tt PSRTIME} and \tempo\footnote{\url{http://tempo.sourceforge.net/}}.

The Arecibo observations were processed using standard {\tt PSRCHIVE} tools.
Using {\tt pat}, TOAs for each of the early Arecibo observations were
 extracted by cross-correlating in the Fourier domain \citep{tay92} 
the folded profile with a noise-free Gaussian template 
derived from the first Arecibo observation. TOA errors were derived using
a Markov chain Monte Carlo method (FDM algorihm in {\tt pat}). These TOAs were then fitted with 
timing solutions using \tempo.
Our timing analysis used the JPL DE405 solar-system ephemeris \citep{sta98b}
and the UTC(NIST) time standard.

In order to obtain the final timing solution presented here a
high signal-to-noise template for each pulsar was constructed.
To construct the templates, the Arecibo subbanded search mode observations were folded with the
best ephemeris. A high signal-to-noise profile was then
constructed by summing the profiles together, weighting by the signal-to-noise
ratio of each observation. The summed profiles were subsequently smoothed using
the {\tt PSRCHIVE} tool {\tt psrsmooth} 
to create a final template profile.
For PSR~J0557+1551, the procedure outlined above was used to construct a template profile, but using the coherent-mode 
data instead.
Figure \ref{fig:profs} shows
the summed profiles and smoothed templates. 
The smoothed templates were used to extract TOAs that were used in the final \tempo\ fit.

For nearly circular binary systems, there is a high correlation between
the longitude of periastron, $\omega$, and the epoch of periastron passage, $T_0$,
in timing models.
Since the four binary pulsars presented here are nearly circular, we use the ELL1 orbital model \citep{lcw+01} in \tempo.
It is parametarized with $\epsilon_1=e\sin\omega$, $\epsilon_2=e\cos\omega$,
and $T_{asc}=T_0-\omega P_b / 2\pi$, where $e$ is the eccentricity, $T_{asc}$ is the epoch 
of the ascending node, and $P_b$ is the orbital period.
This parameterization breaks the covariance but it is an approximation to first order in $e$ and so is
valid only when $xe^2$ is much smaller than the TOA precision. This is true for all the binary MSPs we are timing here.

The TOA uncertainties for each telescope and data-taking mode
(Jodrell Bank, Arecibo incoherent-search mode, and Arecibo coherent mode) were increased 
by a scaling factor (EFAC) that yields a reduced $\chi^2$ equal to unity.
This results in more conservative estimates for the timing parameter uncertainties.
This is standard practice in pulsar timing and it is well known
that formal TOA uncertainties are often underestimated.
These scaling factors (EFACs) are listed in Table \ref{ta:params} along with
our best-fit timing solutions and derived parameters. The residuals are shown in Figure \ref{fig:resids}.
The quoted uncertainties in \ref{ta:params} are 68\% confidence limits as reported by \tempo.

To make a refined measurement of the DM for each pulsar, we split the 600-MHz bandwidth of each PUPPI 
observation into six frequency bands 
and derived six TOAs from each observation for each 100-MHz band. We took our best-fit
solution for each pulsar and held all parameters except for DM fixed. We then fit the TOAs
using all frequency bands to measure a DM. DMs measured in this fashion are listed in
Table \ref{ta:params}.

We also fit for proper motions for PSRs J0557+1551, J1902+0300, and J1905+0453. 
We found that for all three sources the measured values were consistent with no proper motion 
in either right ascension or declination within 95.4\% ($2\sigma$) confidence intervals.
The upper limits on the total proper motion are shown in Table \ref{ta:params}.
The timing baselines for PSRs J1850+0244 and J1943+2210 were not long enough to break
degeneracies with other parameters and place meaningful limits on the proper motion.

\subsection{Flux and Polarization Measurements}

\begin{figure*}
\includegraphics[angle=-90,width=6in]{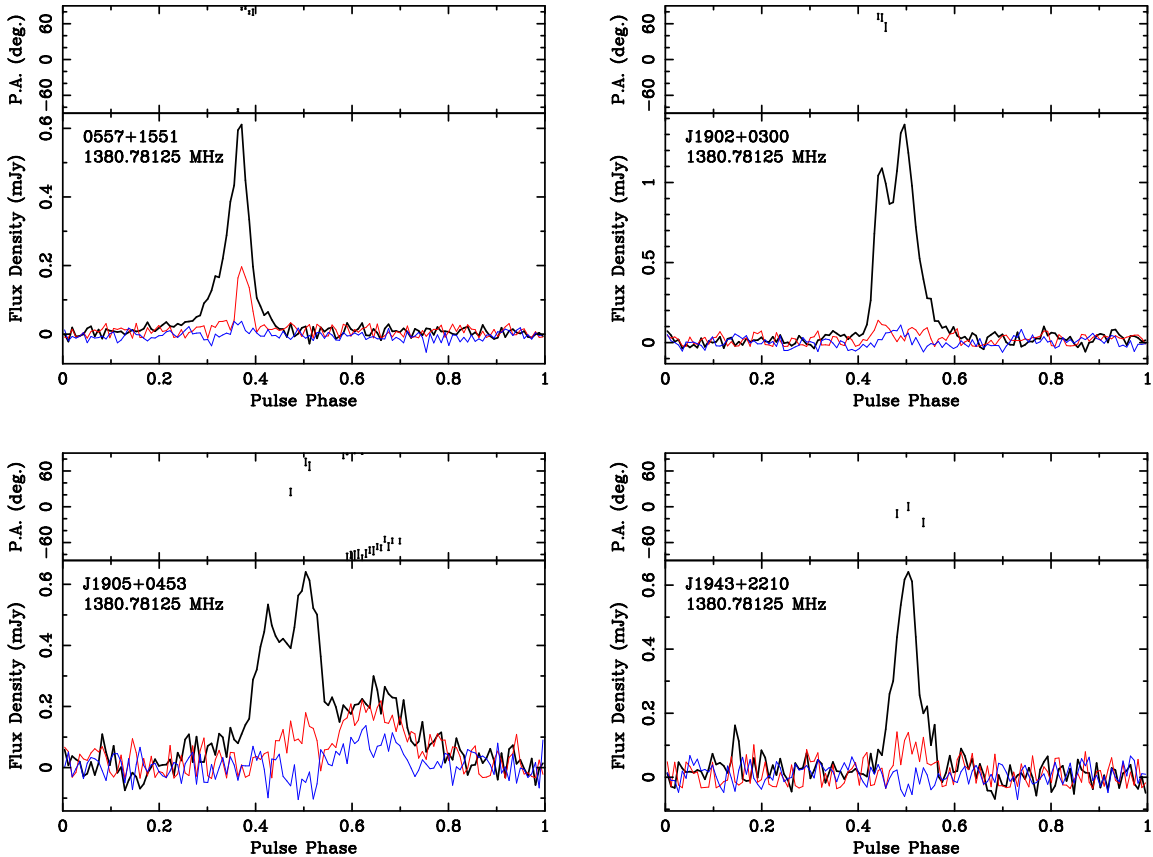}
\figcaption{ Polarization profiles for the four MSPs that had data taken with full Stokes parameters. 
In the bottom panel of each plot, the black, red, and blue lines show the total, linearly polarized, and circularly polarized
flux, respectively. The top panels show the position angle of the linear polarization for profile bins with a linear polarization having
a signal-to-noise greater than three. An RM could only be determined for PSR~J0557+1551 and so its polarization has been 
corrected for Faraday rotation whereas the others have not.
\label{fig:polar}
}
\end{figure*}

\begin{figure}
\plotone{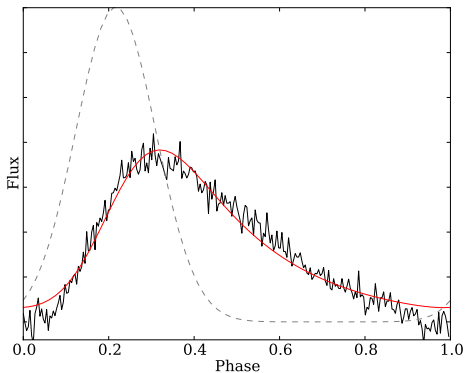}
\figcaption{ Pulse profile of PSR~J1850+0244 (black line) fitted with a wrapped exponentially modified gaussian (red line). The unscattered 
gaussian, i.e. the underlying profile of the pulsar is represented by the dashed grey line. See text for parameters of the fit.
\label{fig:1850prof}
}
\end{figure}

The Arecibo coherent-mode observations were flux- and polarization-calibrated
using observations of a noise diode and of bright quasars (J1413+1509 and B1442+10)
observed for the NANOGrav collaboration \citep{dfg+13}. The calibrations were performed using the 
{\tt PSRCHIVE} tool {\tt pac} and the {\em SingleAxis} model, which assumes that the two polarization receivers 
are perfectly orthogonal.
These calibrated observations provide a full set of Stokes parameters for each 10-s integrated pulse profile 
and so measurements of the flux and polarization characteristics are possible. 
Observations in this mode were made for four of the five pulsars, with PSR~J1850+0244 being the exception
due to its more recent discovery.
The set of calibrated coherent-mode observations were summed and 
Figure \ref{fig:polar} shows the summed polarization profiles for each of the four pulsars.

In order to measure and correct for the effect of Faraday rotation on the polarization of the pulsars,
rotation measures (RMs) were determined by summing the profiles in eight frequency sub-bands
for all the calibrated coherent-mode observations. 
A set of trial RMs was tested by appropriately rotating Stokes $U$ and $Q$ for each frequency channel
and then measuring the total linearly polarized flux, $L$. The RM and its uncertainty were then measured from the 
RM-$L$ curve. Only PSR~J0557+1551 provided a significant measurement with $\mathrm{RM}=26\pm1$\,rad\,m$^{-2}$ 
where the uncertainty is statistical.
This value may contain a contribution from ionospheric Faraday rotation. \citet{hml+06} find for RMs for 223 pulsars at 1.4\,GHz
that the ionospheric RM is typically between $-1$ and $-5$\,rad\,m$^{-2}$. 
So we add in quadrature an uncertainty of $\pm3$\,rad\,m$^{-2}$ to the measurement error to get  $\mathrm{RM}=26\pm3$\,rad\,m$^{-2}$.
The polarization profile of PSR~J0557+1551 in Figure \ref{fig:polar} has been corrected for Faraday rotation using the measured
RM value.

The calibrated observations were also used to measure the flux densities of the four pulsars. 
An off-pulse region was manually selected and its mean was subtracted from the calibrated
profiles. For PSRs~J1902+0300 and J1943+2210, the phase range excluding $\pm15\%$ of pulse phase
from the peak was used as the off-pulse region. 
For J0557+1551, due to its narrower pulse, $\pm10\%$ from the peak was excluded.
For J1905+0453, 30\% of phase preceeding the peak and 40\% following the peak was excluded.
The mean flux density of the baseline-subtracted profile was then computed.
Those values are presented in Table \ref{ta:params}.
Note that the quoted uncertainties are only statistical and do not take into account any
uncertainties from the calibration procedure.

\section{Discussion}
\label{sec:disc}

\begin{figure*}
\plottwo{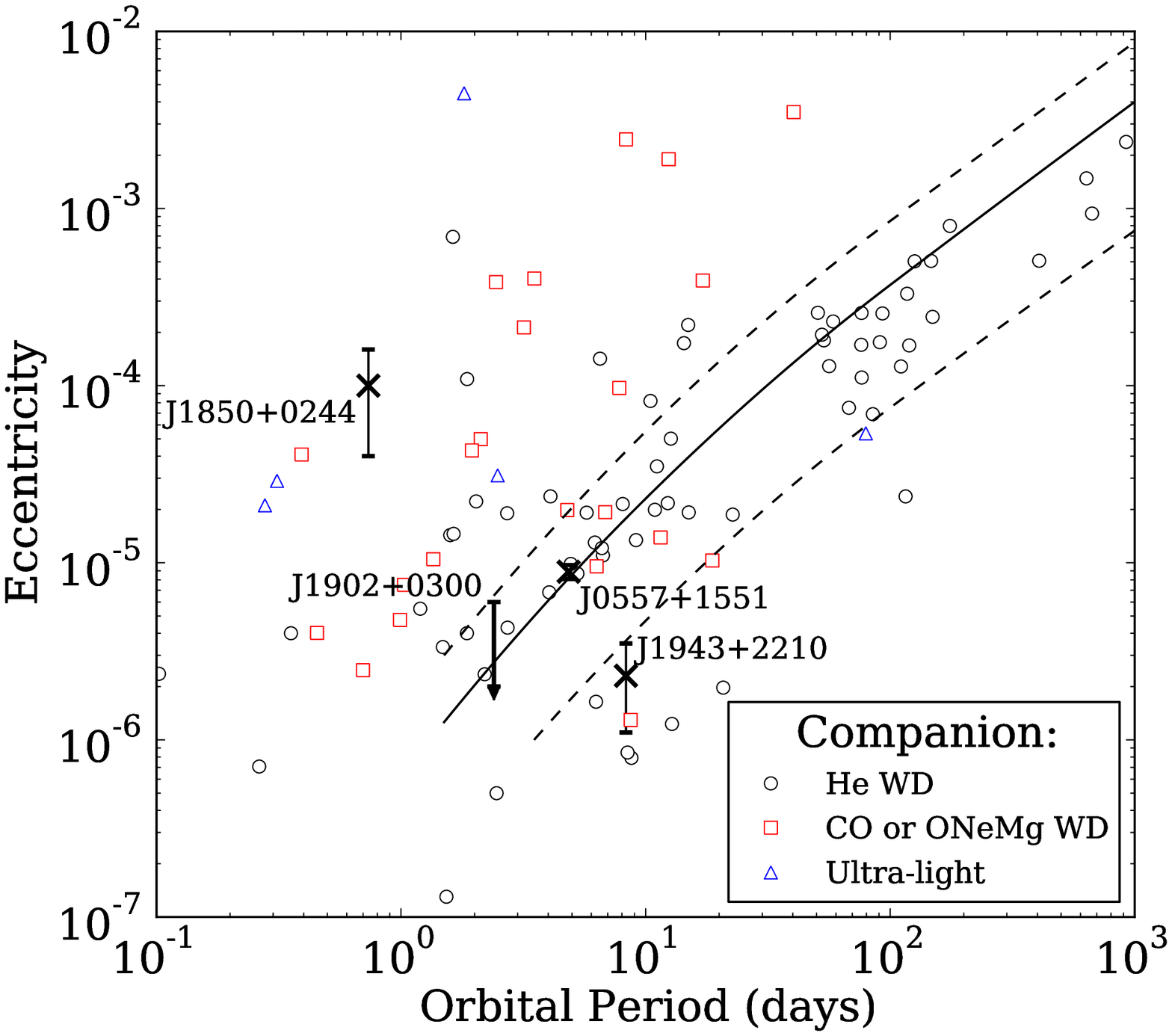}{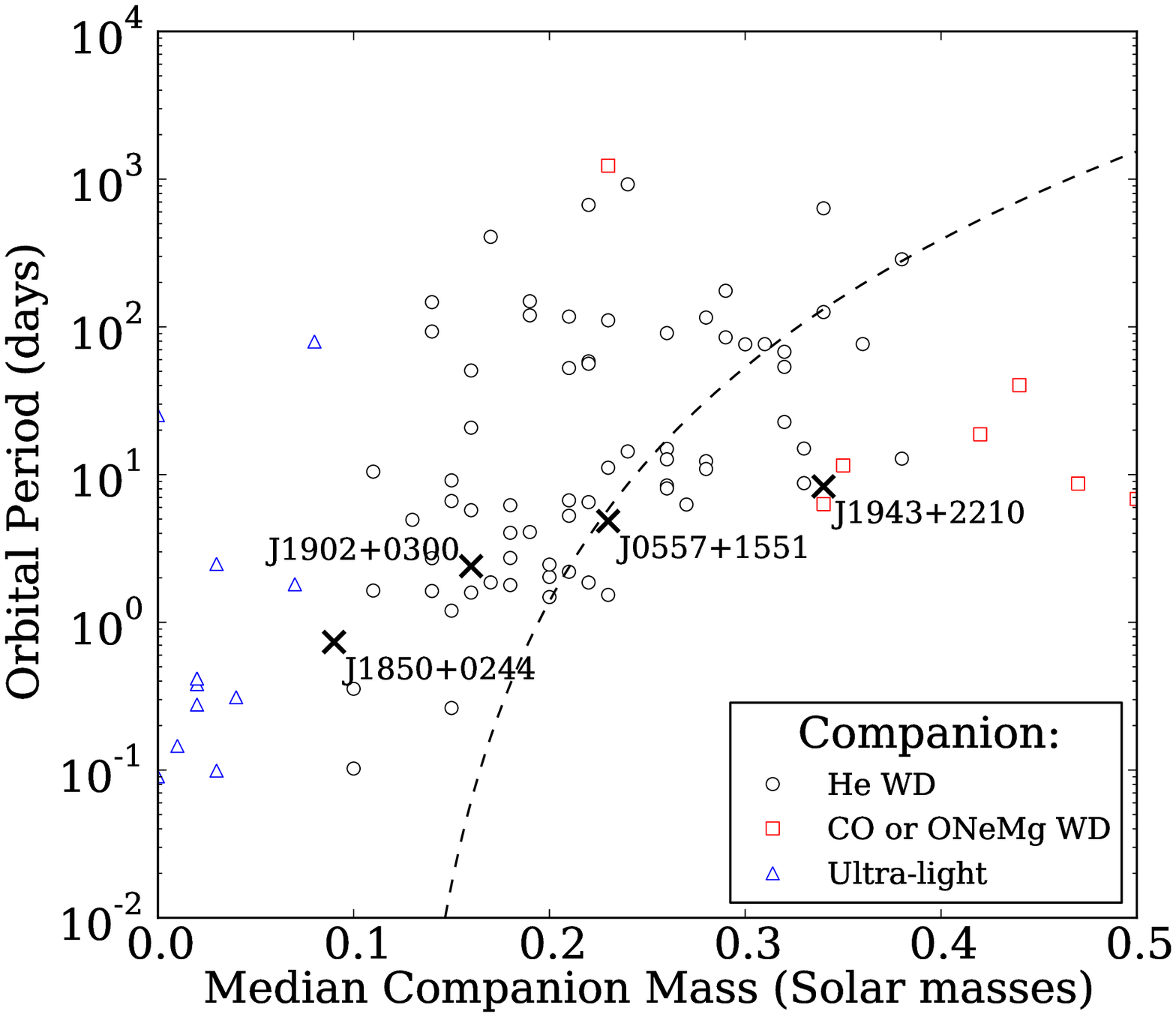}
\figcaption{ {\em Left:} Eccentricity as a function of orbital period for Galactic binary pulsars from the ATNF pulsar catalog.
The solid line shows the relation from the HeWD-MSP binary formation model of \citet{pk94} and the dotted lines show the 95\% confidence intervals
for that model. Crosses with error bars show the values and uncertainies at a 95\% confidence level 
for the binary systems in this work. As the measured eccentricity
for PSR~J1902+0300 is consistent with zero it is shown as an upper limit.
{\em Right:} Companion mass versus orbital period for Galactic field binary pulsars from the ATNF pulsar catalog. The dotted line
shows the relation predicted by \citet{ts99} for MSP-WD systems formed from LMXBs.
Crosses show the four binary systems presented in this work. Other symbols represent the type of companion as tabulated
in the ATNF pulsar catalog.
\label{fig:phinney}
}
\end{figure*}

We have reported the discovery of five MSPs from the PALFA survey. 
We presented phase-coherent timing solutions of the five pulsars using observations from the Arecibo and Lovell telescopes. 
One of the pulsars was found to be isolated, and the rest are in nearly circular binaries.
The DMs of the MSPs are high (see Figure \ref{fig:pdm}), especially PSR~J1850+0244 ($\mathrm{DM}=540$\,pc\,cm$^{-3}$), 
which is the highest DM MSP yet discovered.
PALFA has to date discovered 18 MSPs. All but one, and all five presented in this work, have DMs greater than 100\,pc\,cm$^{-3}$.
This range is in the top 15\% of non-PALFA Galactic MSPs and the addition of the PALFA discovered MSPs represents a 70\% increase
in the number of MSPs with DM\,$>100$\,pc\,cm$^{-3}$.
Fourteen of the PALFA-discovered MSPs have DM/$P$ ratios larger than 30\,pc\,cm$^{-3}$\,ms$^{-1}$, compared to 16 non-PALFA Galactic MSPs.
PALFA discoveries have thus almost doubled the number of known Galactic MSPs with DM/$P$ ratios larger than this value (Figure \ref{fig:pdm}).
Here we discuss some of the notable properties of each of the five MSPs.

\subsection{PSR~J0557+1551}

PSR~J0557+1551 is the most rapidly rotating of the five MSPs with a period of 2.55\,ms. 
It is in a 4.9-day binary orbit and it has a DM of 103\,pc\,cm$^{-3}$. Its DM implies a distance, estimated from the NE2001 model of 
Galactic free-electron content \citep{cl02},
of 2.9\,kpc. The minimum companion mass (i.e. the mass implied by the measured mass function assuming
an orbital inclination angle of $i=90\degrees$ and a pulsar mass of 1.4\,$M_\odot$) 
is 0.20\,$M_\odot$ and the median companion mass (i.e. $i=60\degrees$) is 0.23\,$M_\odot$.
Our timing solution has an rms residual of 4.9\,$\mu$s and spans 2.4\,years.
PSR~J0557+1551 is located in the Galactic anticenter
and so is in a region with a relatively low density of known pulsars. Since its rms residuals and
TOA errors are close to those in existing pulsar timing arrays \citep[e.g.][]{dfg+13}, along with the fact that it is located in an area 
where pulsar timing arrays presently
have low sensitivity \citep[e.g.][]{abb+14}, it may be a good candidate for use in gravitational wave detection.

\subsection{PSR~J1850+0244}

PSR~J1850+0244 is the highest DM MSP yet discovered (540\,pc\,cm$^{-3}$); its DM-derived distance is 10.4\,kpc, the furthest for any known Galactic MSP. 
It has a spin period of 4.48\,ms and is in a 0.74-day orbit with a companion with minimum and median masses of 0.07\,$M_\odot$ and 0.09\,$M_\odot$,
respectively. 
The timing solution for this pulsar has an rms residual of 57\,$\mu$s and spans 10\,months.

The profile of PSR~J1850+0244 shows a clear scattering tail. It also shows a slow rise implying that the underlying
pre-scattered profile is quite broad. We fit the profile with a wrapped exponentially modified Gaussian, i.e. an underlying
Gaussian profile convolved with an exponential decay, caused by the scattering, that is wrapped at the period of the pulsar. 
We show the pulse profile
in Figure \ref{fig:1850prof} along with the resulting best-fit exponentially modified Gaussian model.
The best-fit exponential decay timescale was found to be $1.02\pm 0.04$\,ms; this is close to the value predicted
for scattering at 1.4\,GHz by the NE2001 model of 1.3\,ms. The best-fit width parameter of the underlying Gaussian
is $\sigma=0.41\pm0.02$\,ms which corresponds to a width at 50\% of the peak of $W_{50}=0.97$\,ms (a $\sim20$\% duty cycle).
We caution that the exponentially-convolved Gaussian fitting function is only an approximation to the scattering of the actual pulse shape \citep{bcc+04}, 
and that the above quoted uncertainties in the scattering time and intrinsic pulse width are solely statistical, and do not take into account
the likely dominant systematic uncertainties from such an approximation.

\subsection{PSR~J1905+0453}

PSR~J1905+0453 is an isolated pulsar with a spin period of 6.09\,ms, a DM of 183\,pc\,cm$^{-3}$, and a DM-implied distance of 4.7\,kpc.
Its timing solution has an rms residual of 33\,$\mu$s and spans 2.1\,years.

PSR~J1905+0453 has the second longest period out of all known isolated MSPs (PSR~J1730$-$2304, with a spin period of 8.12\,ms has the longest period among
isolated MSPs; \citealp{lnl+95}).
It also has the lowest $\dot{E}$ out of any known isolated MSP ($1\times10^{33}$\,erg\,s$^{-1}$). 
Isolated recycled pulsars are thought to originate from disruption
due to a second supernova \citep{lma+04}, ejection from a triple system \citep{fbw+11}, or ablation of the companion by the pulsar wind \citep{fbb+90,bbb+11}. 
The short spin period requires a companion lifetime that is much longer than that of stars massive enough to explode as supernovae.
Ablation requires enough spin-down luminosity to 
power a pulsar wind that can disintegrate its companion. The question is then whether or not the low $\dot{E}$ 
of PSR~J1905+0453 would have been sufficiently large to power ablation earlier in its life as an MSP. 

The $\dot{P}$, and thus $\dot{E}$, of PSR~J1905+0453 will be biased by the acceleration in the Galactic plane. Using the formulae from \citet{nt95},
we find that the motion of PSR~J1905+0453 in the Galactic plane causes an effective $\dot{P}$ of $1.3\times10^{-21}$ and that the
acceleration in the direction perpendicular to the plane is negligible. The corrected $\dot{P}$ and $\dot{E}$ are thus
$4.0\times10^{-21}$ and $7.0\times10^{32}$\,erg\,s$^{-1}$.

The $\dot{P}$ of PSR~J1905+0453 could also be influenced by the Shklovskii effect whereby the space velocity relative
to the observer causes an apparent contribution to $\dot{P}$. Since the proper motion measurement for PSR~J1905+0453 is only
an upper limit, the contribution from the Shklovskii effect cannot be calculated. However, if we assume that the observed $\dot{P}$
is caused exclusively by the Shklovskii effect, it implies a transverse velocity of 200\,km/s. This is a reasonable
velocity; according to the ATNF pulsar catalog, $\sim40\%$ of pulsars 
with measured proper motions have inferred transverse velocities equal or greater than 200\,km/s. 
The true $\dot{P}$, and hence $\dot{E}$, of PSR~J1905+0453 could therefore be much lower than the value in Table \ref{ta:params}.

For now, let us assume that the contribution from the Shklovskii effect is minimal.
The spin-down luminosity of PSR~J1905+0453 was of course higher in the past. If we assume pure magnetic-dipole braking,
a constant magnetic field,
and that the MSP was recycled at most 14\,Gyr ago (the age of the Universe), the fastest period and highest $\dot{E}$
that PSR~J1905+0453 could have had at formation would be 3.34\,ms and $8\times10^{33}$\,erg\,s$^{-1}$, respectively.
This spin-down luminosity is comparable to those of several black widow pulsars, 
the least energetic of which have $\dot{E}\sim1\times10^{34}$\,erg\,s$^{-1}$ \citep{mr13}.
Since black widow pulsars are thought to be in the process of ablating their companion, this suggests that it may have been
possible for PSR~J1905+0453 to have done so given its spin-down power 
if it originated in a binary similar to those of known black widow pulsars and the contribution of the Shklovskii effect 
to its measured $\dot{E}$ is not dominant.

\subsection{PSR~J1902+0300}

PSR~J1902+0300 has a period of 7.80\,ms, a DM of 254\,pc\,cm$^{-3}$, and a NE2001 implied distance of 5.9\,kpc. 
It is in a 2.4-day orbit and its companion has a minimum mass of 0.14\,$M_\odot$ and
a median mass of 0.16\,$M_\odot$.
Our timing solution has an rms residual of 14\,$\mu$s and spans 2.3\,years.

\subsection{PSR~J1943+2210}

PSR~J1943+2210 has a period of 8.31\,ms, a DM of 174\,pc\,cm$^{-3}$, and a NE2001 implied distance of 6.2\,kpc. 
It is in a 8.3-day orbit with a minimum companion mass of 0.28\,$M_\odot$ and a medium mass of 0.34\,$M_\odot$.
Its timing solution has an rms residual of 13\,$\mu$s and spans 1.2\,years.

\subsection{Counterparts in other wavelengths}

Because the pulsars presented here are quite faint and distant we would probably not expect
 counterparts in other wavelengths to be easily detected. 
The brightest white dwarfs have optical absolute magnitudes of $\sim10$ \citep{co07}. The closest of the five binary systems in this work,
according to their DM-derived distances, is PSR~J0557+1551 at 3\,kpc. At this distance a $M_V=10$ white dwarf would have an
apparent magnitude of $\sim22$. This is quite faint and it is likely that the actual companion is fainter than this. 

To see whether we expect to see gamma-ray counterparts for the MSPs, we can compare the spin-down energy incident at the Earth, $\dot{E}/D^2$
of the pulsars here with those of the {\em Fermi} detected MSPs. According to the ATNF pulsar catalog, most {\em Fermi}-detected MSPs have $\dot{E}/D^2>10^{33}$\,erg\,s$^{-1}$\,kpc$^{-2}$. 
Only PSR~J0557+1551 has a $\dot{E}/D^2$ higher than this, and it is possible that a signifcant portion of the measured $\dot{P}$ is due to the Shklovskii effect.
Furthermore, the other four MSPs are located in the inner Galactic plane where the gamma-ray background is high.

Indeed, we find no coincident optical, infrared, X-ray or $\gamma$-ray
sources in the SIMBAD\footnote{\url{http://simbad.u-strasbg.fr/simbad/}} or 
HEASARC\footnote{\url{http://heasarc.gsfc.nasa.gov/}} databases or in the {\em Fermi} LAT 2-year Point Source Catalog \citep{naa+12}
for any of the MSPs reported here. 

\subsection{Nature of the Binary Companions}

None of the four binaries presented here show any evidence for eclipses or matter in the system in their timing residuals, 
which suggests that none of the companions are likely close to filling their Roche lobes.
Given the orbital separation and masses of the components of a binary system, the size of the Roche lobe of the companion can be calculated \citep{tv06}.
If we assume a typical inclination of $60\degrees$ and a pulsar mass of 1.4\,$M_\odot$, the sizes of the companion Roche lobes for
all four binary systems can accomodate both white dwarfs and main sequence stars.  

In Figure \ref{fig:phinney} we plot the binary parameters of the four binary MSPs presented here 
along with those from all Galactic-field binary MSPs, taken from the ATNF pulsar catalog. 
The plots of orbital period versus eccentricity and companion mass versus orbital period allow
us to compare the binary parameters to the predicted relations from \citet{pk94} and \citet{ts99}
for helium white-dwarf (HeWD)-MSP binaries that evolved through long-term, stable, mass transfer
from a red giant progenitor onto a neutron star in a low-mass X-ray binary (LMXB).

The median companion masses implied by the mass functions of PSR~J1902+0300 and PSR~J0557+1551 are in the range expected for HeWD companions \citep{tlk12}.
They also fall on the orbital period-eccentricity relation predicted by \citet{pk94} and near the companion mass-orbital period relation predicted by
\citet{ts99}. Thus, the likely nature of their companions are HeWDs. 

The other two binary systems fall slightly outside of the parameters expected for a system with a HeWD companion
and so formation scenarios other than LMXB evolution into a HeWD may
have occured. PSR~J1943+2210 has a companion mass that is higher than expected for a HeWD and 
falls outside of the relation from \citet{pk94} with a lower
than expected eccentricity. The high companion mass implies that its companion could be a carbon-oxygen white dwarf.
It would therefore be considered an intermediate-mass binary pulsar \citep[IMBP;][]{cam96c} a class of currently about 20 objects \citep{tlk12}.

PSR~J1850+0244, on the other hand, has a lower mass than expected for a HeWD and a higher than predicted eccentricity. 
In the $M_c-P_b$ plot of Figure \ref{fig:phinney}, it falls close to the systems classified as ``Ultra-light'', 
many of which are considered black widows \citep{mr13}, raising the possibility
that it has interacted with its companion through ablation and accretion processes, as the black-widow systems are thought to have done.
Of course, the true companion mass of PSR~J1850+0244 could be higher if the system has a low inclination. The relativistic MSP-HeWD binaries PSRs~J0348+0432
and J1738+0333, which like PSR~J1850+0244 have orbital periods less than a day and median companion masses $\sim0.1\,M_\odot$,
have measured true companion masses determined from their post-Keplerian parameters and white dwarf spectroscopy. 
PSR~J0348+0432 has a companion mass of 0.172\,$M_\odot$ and an inclination of $40\degrees$ \citep{afw+13}. PSR~J1738+0333 has a companion mass 
of 0.181\,$M_\odot$ and an inclination of $32\degrees$  \citep{akk+12}. 
These companion masses are close to those predicted for HeWDs by \citet{ts99}.  
If we assume that PSR~J1850+0244 has a HeWD companion with a mass of 0.19\,$M_\odot$ as predicted by \citet{ts99},
the inclination of the binary system would be $24\degrees$ for a pulsar mass of 1.4\,$M_\odot$ and $30\degrees$ for a pulsar mass of 2.0\,$M_\odot$.
This is not an unreasonably fortuitous orientation and so it is plausible that PSR~J1850+0244 has a low inclination and a standard HeWD companion.

\subsection{Future prospects for mass measurements}

\begin{figure*}
\plottwo{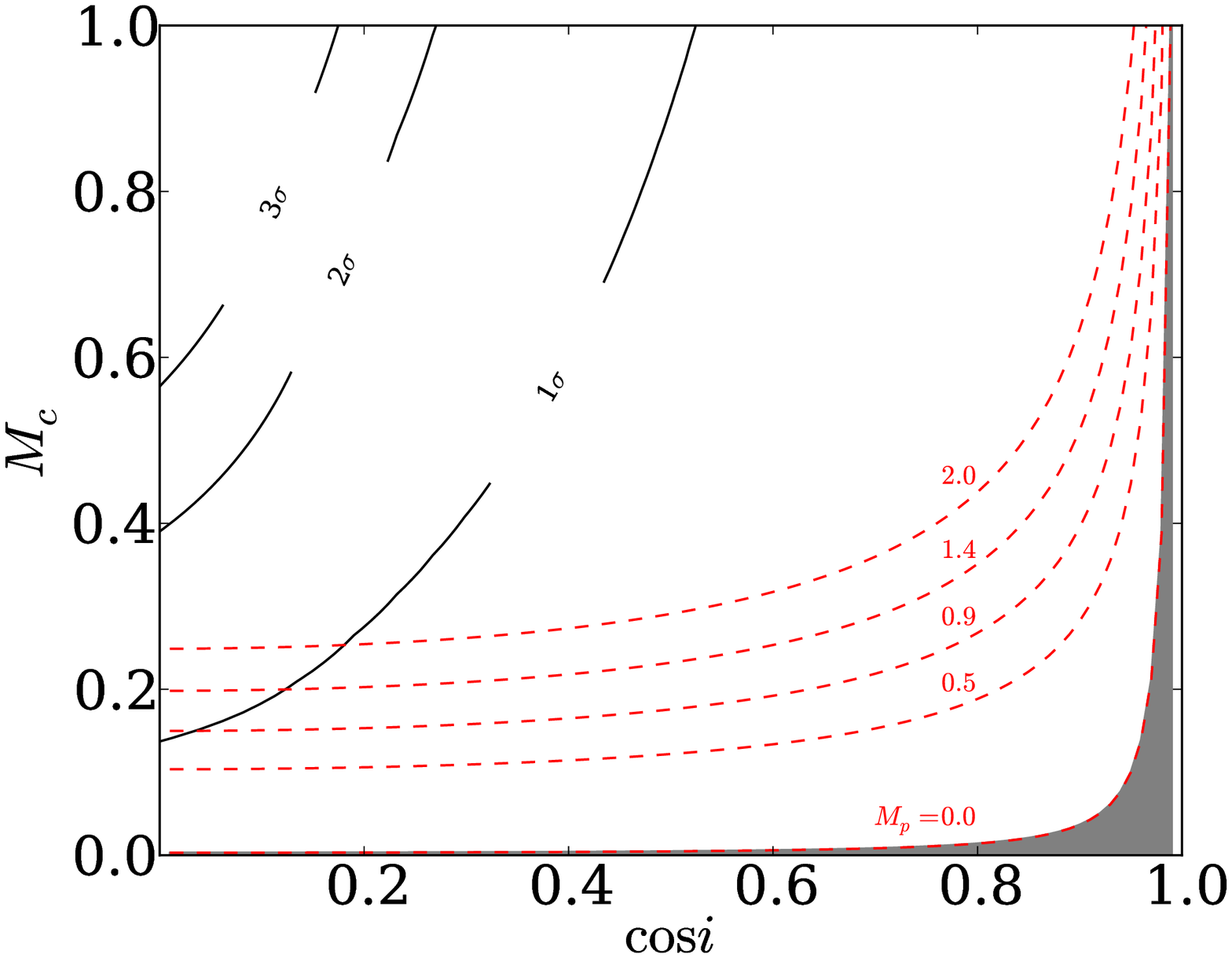}{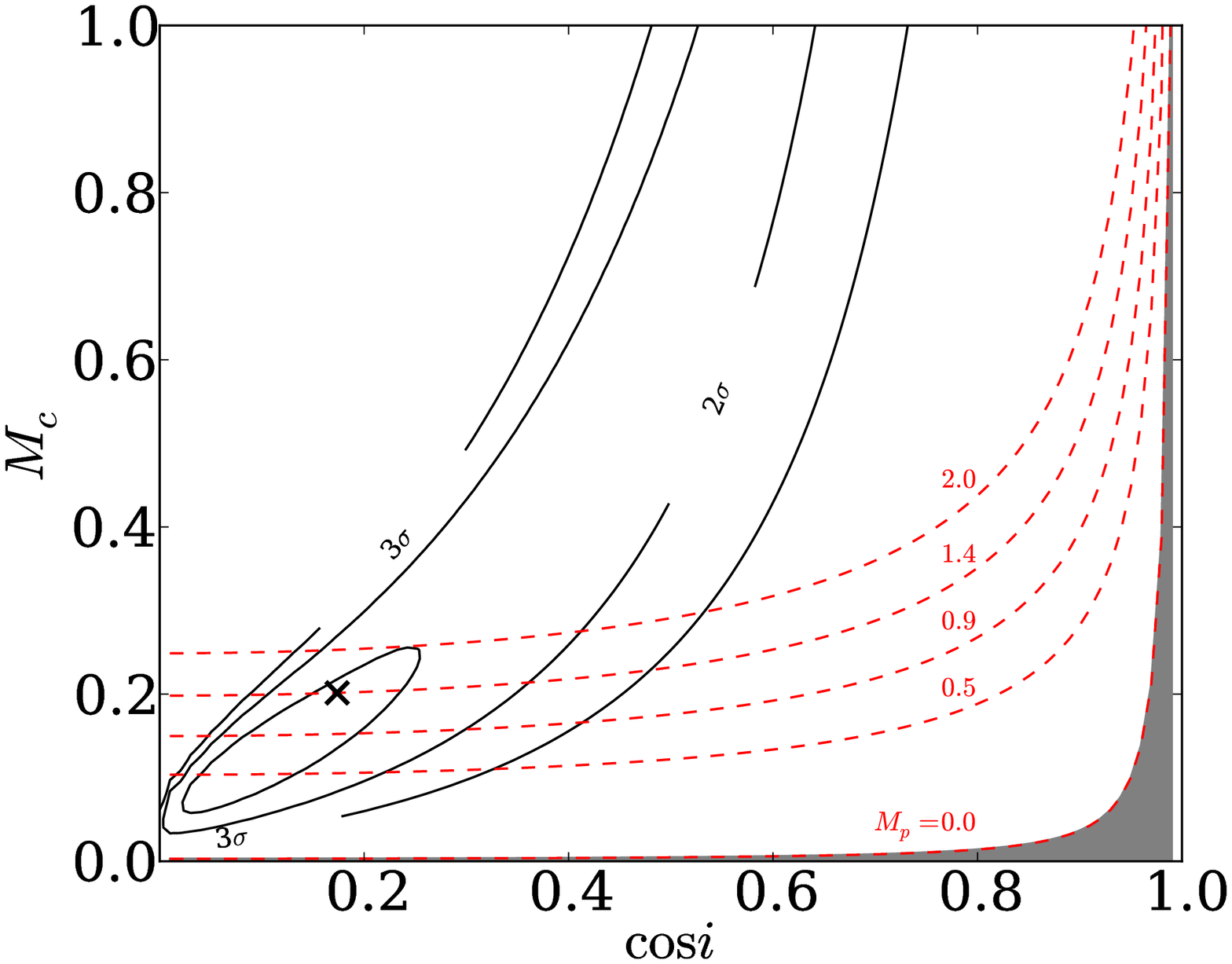}
\figcaption{ Maps of companion-mass and inclination angle for PSR~J0557+1551. The left plot shows binary configurations allowed by current timing data and the right
shows configurations for a simulated
data set with $M_p=1.35M_\odot$ and $i=80\degrees$ (see text). 
The black cross denotes the simulated binary configuration.
The black contours enclose 68.3\% (1$\sigma$), 95.4\% (2$\sigma$), and 99.7\% (3$\sigma$)
of allowed configurations. The red dashed lines denote the pulsar mass for the configuration.
\label{fig:shap}
}
\end{figure*}

Masses of the components in a MSP binary can be measured using the Shapiro delay effect \citep[e.g.][]{rt91a,sns+05,dpr+10}.
The Shapiro delay is parameterized in timing models by the post-Keplerian parameters $r$ and $s$. 
The Shapiro range, $r$, and shape, $s$, parameters are equal to the companion mass in units of light seconds and the sine of the inclination, respectively.

In the ELL1 model, when binary systems have measured $\epsilon_1=e\sin\omega$ greater than zero and $\epsilon_2=e\cos\omega$ consistent with zero,
it may be the case that the eccentricity, $e$, is actaully zero and the non-zero $\epsilon_1$ is due to a Shapiro delay signal \citep{lcw+01}.
For two of the four binary MSPs in this work, PSRs~J0557+1551 and J1850+0244, $\epsilon_1>0$ and $\epsilon_2=0$. 
For these systems, given an assumed inclination, we can calculate the companion mass (i.e. Shapiro $r$)
implied by the measured $\epsilon_1$ assuming that the true eccentricity is consistent with zero \citep[see Appendix A2 of][]{lcw+01}.
If the implied companion masses are in a reasonable range given the measured mass functions, this would give evidence that the measured $\epsilon_1$ is due to Shapiro delay.
For PSR~J0557+1551, the lowest implied companion mass (i.e. at high inclination) is $\sim2\,M_\odot$ and for PSR~J1850+0244 it is $\sim2.5\,M_\odot$.
Since the measured mass functions for these two MSPs would require unreasonably high pulsar masses in order for their companions to be so heavy, this implies that
the measured values of $\epsilon_1$ for PSRs~J0557+1551 and J1850+0244 are not dominated by Shapiro delay signals.

In order to test for a Shapiro delay signal in each of the binary pulsars, a timing model was fit using the \tempo\ DD model \citep{dd85,dd86}
which includes post-Keplerian parameters in the timing model. 
We fit the TOAs using \tempo\ with the best-fit parameters listed in Table \ref{ta:params} with the addition of a test $M_c$ and $\cos i$.
We searched a grid of $M_c$ and $\cos i$ ranging from $0.0 \le M_c \le 1.0$ and $0.0 \le \cos i \le 1.0$. 
For all but PSR~J0557+1551, all points in the grid provided acceptable fits at greater than a 95.4\% level and so no limits could be placed.
For PSR~J0557+1551, the left panel of Figure \ref{fig:shap} shows the upper-limit contours placed on $M_2$ and $\cos i$. 
It is clear that a small portion of reasonable binary system parameters are inconsistent with the data at a 1$\sigma$ confidence level, namely at high inclination and
high pulsar mass. 

These current limits are not very constraining, but future observations may yield interesting results. To investigate this,
we constructed a simulated set of TOAs with uncertainties and rms residuals of approximately 1\,$\mu$s. The simulated timing campaign spanned a year
with a one month cadence and included ten 2.5-h observations that each occured at one of the ten conjunctions of the system that will be visible
at Arecibo in that time span. The right panel of Figure \ref{fig:shap} shows the limits that could be placed with such a campaign given
 a pulsar mass of 1.4$M_\odot$ and an inclination of 80$\degrees$ which imply a companion mass of 0.20$M_\odot$.
Note that if PSR~J0557+1551 has a more favorable binary configuration (i.e. higher component masses or inclination) or if the timing precision
could be improved to better than 1-$\mu$s then more stringent measurements could be made with a campaign similar to that simulated.

\section{Conclusions}

The search for MSPs is heavily motivated by the unique and exciting studies that are enabled
 by finding more systems that provide laboratories for such science. 
PALFA provides a unique window on the MSP population, discovering
more distant and fainter MSPs than any other past or current survey.
We have presented here the discovery and follow-up timing of
five MSPs discovered in the PALFA survey as well as discussed their interesting properties.
One MSP, PSR~J1905+0453 is isolated and has the lowest spin-down luminosity yet measured for an isolated MSP.
Three of the binary MSPs likely have HeWD companions and one, PSR~J1943+2210, may have a carbon-oxygen WD companion.
At least one of the five, PSR~J0557+1551, has the potential to allow mass measurements and may be useful
in pulsar timing arrays. 
PSR~J1850+0244 is the highest DM MSP yet discovered with a DM of 540.
Further discoveries in the ongoing PALFA survey will expand our knowledge
of the MSP population deep into the Galactic plane as well as help constrain fundamental physics and astrophysics.

We thank the anonymous referee for their constructive comments.
The Arecibo Observatory is operated by SRI International under a cooperative agreement with the National Science Foundation (AST-1100968), 
and in alliance with Ana G. M\'endez-Universidad Metropolitana, and the Universities Space Research Association. 
P.S. acknowledges support from an NSERC Alexander Graham Bell Canada Graduate Scholarship.
V.M.K. acknowledges support from an NSERC Discovery Grant and Accelerator Supplement, the FQRNT Centre de Recherche en Astrophysique du Qu\'ebec, 
an R. Howard Webster Foundation Fellowship from the Canadian Institute for Advanced Research (CIFAR), the Canada Research Chairs Program and the Lorne
Trottier Chair in Astrophysics and Cosmology. 
J.W.T.H. acknowledges funding from an NWO Vidi fellowship and ERC Starting Grant ``DRAGNET'' (337062).
P.C.C.F. and L.G.S.
gratefully acknowledge financial support by the European
Research Council for the ERC Starting Grant BEACON under
contract no. 279702.
P.L. acknowledges the support of IMPRS Bonn/Cologne and FQRNT B2.
Work at Cornell University was supported in part by the National Science Foundation  (PHYS-PHY‐1104617). 
Pulsar research at UBC is supported by an NSERC Discovery Grant and Discovery Accelerator Supplement, and by the Canadian Institute for Advanced Research.
Computations were made on the supercomputer Guillimin from McGill University, 
managed by Calcul Qu\'ebec and Compute Canada. The operation of this supercomputer is funded by the 
Canada Foundation for Innovation (CFI), NanoQu\'ebec, RMGA and the Fonds de recherche du Qu\'ebec - Nature et technologies (FRQ-NT).

\bibliographystyle{apj}
\bibliography{journals_apj,/homes/borgii/pscholz/Documents/papers/myrefs,modrefs,psrrefs,crossrefs,/homes/janeway/maggie/Tex/maggie_refs}
%\bibliography{journals_apj,myrefs,modrefs,psrrefs,crossrefs,/homes/janeway/maggie/Tex/maggie_refs}
%\bibliography{journals1,myrefs,modrefs,psrrefs,crossrefs}

%\includepdf[pages={1},lastpage=1,landscape=true,turn=false]{table.pdf}

\clearpage
\begin{landscape}
%\begin{turnpage}
\begin{deluxetable*}{lccccc}
\tabletypesize{\footnotesize}
\tablewidth{0pt}
\tablecolumns{6}

\tablecaption{Timing Parameters for five MSPs
\label{ta:params}}
\tablehead{
\colhead{PSR~J}&\colhead{0557+1551}&\colhead{1850+0244}&\colhead{1902+0300}&\colhead{1905+0453}&\colhead{1943+2210}
}

\startdata
\tableline
\multicolumn{6}{c}{Measured Parameters} \\
\tableline
Right ascension (J2000)                             & $05:57:31.44918(9)$               & $18:50:41.683(1)$                & $19:01:59.6168(1)$               & $19:04:59.3848(2)$               & $19:43:16.4910(1)$      \\
Declination (J2000)                                 & $15:50:06.046(8)$                 & $02:42:57.44(3)$                 & $03:00:23.324(5)$                & $04:51:54.953(8)$                & $22:10:23.151(2)$       \\
Total proper motion, $\mu_T$ (mas\,yr$^{-1}$)$^a$   & $<80$                             & -                                & $<77$                            & $<96$                            & -                       \\    
Spin frequency, $\nu$ (s$^{-1}$)                    & $391.18012787837(8)$              & $223.221059721(3)$               & $128.25813729784(5)$             & $164.1397912192(1)$              & $196.6887956594(2)$     \\ 
Frequency derivative, $\dot{\nu}$ (s$^{-2}$)        & $-1.125(3)\times10^{-15}$         & $-8.1(2)\times10^{-15}$          & $-7.52(2)\times10^{-16}$         & $-1.54(6)\times10^{-16}$         & $-3.4(1)\times10^{-16}$ \\
Dispersion measure, DM (pc\,cm$^{-3}$)              & 102.5723(5)                       & 540.068(9)                       & 253.887(2)                       & 182.705(4)                       & 174.089(3)              \\
Orbital period, $P_b$ (d)                           & $4.846550440(4)$                  & $0.73543015(1)$                  & $2.399218509(6)$                 & -                                & $8.31148080(2)$         \\
Projected semimajor axis, $x=a\sin i /c$ (s)        & $4.0544597(8)$                    & $0.452355(9)$                    & $1.842553(1)$                    & -                                & $8.064088(2)$           \\
Time of ascending node, $T_{asc}$ (MJD)             & $56349.2459263(2)$                & $56468.179498(6)$                & $55989.555577(1)$                & -                                & $56392.7364900(7)$      \\
$\epsilon_1 = e\sin\omega$                          & $9.3(4)\times10^{-6}$             & $9(3)\times10^{-5}$              & $3(2)\times10^{-6}$              & -                                & $2.5(8)\times10^{-6}$   \\
$\epsilon_2 = e\cos\omega$                          & $-9(60)\times10^{-8}$             & $-4(40)\times10^{-6}$            & $4(10)\times10^{-7}$             & -                                & $-1.4(4)\times10^{-6}$  \\
\tableline
\multicolumn{6}{c}{Derived Parameters} \\
\tableline
Galactic longitude, $l$ (degrees)                   & 192.68                            & 35.26                            & 36.81                            & 38.81                            & 58.44                            \\        
Galactic latitude, $b$ (degrees)                    & -4.31                             & 1.40                             & -0.97                            & -0.79                            & -0.73                            \\        
Period, $P$ (ms)                                    & $2.5563670767830(5)$              & $4.47986404710(6)$               & $7.796776259723(3)$              & $6.092367929629(5)$              & $5.084173689952(5)$              \\        
Period derivative, $\dot{P}$                        & $7.35(2)\times10^{-21}$           & $1.62(4)\times10^{-19}$          & $4.57(1)\times10^{-20}$          & $5.7(2)\times10^{-21}$           & $8.7(3)\times10^{-21}$           \\
Eccentricity, $e$                                   & $9.3(4)\times10^{-6}$             & $9(3)\times10^{-5}$              & $3(2)\times10^{-6}$              & -                                & $2.9(7)\times10^{-6}$            \\
Longitude of periastron, $\omega$ (deg)             & $91(4)$                           & $90(20)$                         & $80(30)$                         & -                                & $120(10)\times10^{2}$            \\
Time of periastron passage, $T_0$ (MJD)             & $56350.47(5)$                     & $56468.37(5)$                    & $55990.1(2)$                     & -                                & $56395.5(2)$                     \\
Mass function ($M_\odot$)                           & $3.046605(2)\times10^{-3}$        & $1.8375(1)\times10^{-4}$         & $1.166818(2)\times10^{-3}$       & -                                & $8.150639(6)\times10^{-3}$       \\
Minimum companion mass ($M_\odot$)$^b$              & 0.20                              & 0.07                             & 0.14                             & -                                & 0.28                             \\
Median companion mass ($M_\odot$)$^c$               & 0.23                              & 0.09                             & 0.16                             & -                                & 0.34                             \\
Surface dipolar magnetic field, $B$ (G)             & $1.388(2)\times10^{8}$            & $8.6(1)\times10^{8}$             & $6.040(8)\times10^{8}$           & $1.89(3)\times10^{8}$            & $2.13(3)\times10^{8}$            \\
Spin-down luminosity, $\dot{E}$ (erg\,s$^{-1}$)     & $1.738(5)\times10^{34}$           & $7.1(2)\times10^{34}$            & $3.81(1)\times10^{33}$           & $1.00(4)\times10^{33}$           & $2.61(8)\times10^{33}$           \\
Characteristic age, $\tau_c$ (yr)                   & $5.51(2)\times10^{9}$             & $4.4(1)\times10^{8}$             & $2.705(7)\times10^{9}$           & $1.69(6)\times10^{10}$           & $9.3(3)\times10^{9}$             \\
DM distance, $D$ (kpc)$^d$                          &  2.9                              & 10.4                             & 5.9                              & 4.7                              & 6.2 \\
Transverse velocity, $V_T$ (km\,s$^{-1}$)$^a$       & $<1100$                           & -                                & $<2100$                          & $<2100$                          & -   \\
\tableline                                                
\multicolumn{6}{c}{Observation Parameters} \\
\tableline
1400\,MHz mean flux density, $S_{1400}$ (mJy)       & 0.050(6)                          & -                                & 0.113(4)                         & 0.117(9)                         & 0.04(1) \\
1400\,MHz mean radio luminosity, $L_{1400}$ (mJy kpc$^2$)$^e$ & $\sim0.4$               & -                                & $\sim3.9$                        & $\sim2.6$                        & $\sim1.5$ \\
Discovery observation date (MJD)                    & 55907                             & 56466                            & 55863                            & 55817                            & 56389   \\
Timing epoch (MJD)                                  & 56346.00                          & 56554.29                         & 56242.00                         & 56242.00                         & 56404.00\\
Start epoch (MJD)                                   & 55963.08                          & 56546.71                         & 55989.20                         & 56067.92                         & 56391.07\\
Finish epoch (MJD)                                  & 56841.65                          & 56856.14                         & 56828.22                         & 56831.16                         & 56828.25\\
Number of TOAs                                      & 162                               & 160                              & 270                              & 278                              & 147     \\
Typical Integration time per Arecibo TOA (min)      & 10                                & 2                                & 2                                & 2                                & 2       \\
Typical TOA uncertainty ($\mu$s) (Jodrell/AO-search/AO-fold) &  23/7/3                  & 130/50/-                          & 80/15/15                        & 160/30/25                        & 60/10/10\\
TOA error scale factor, EFAC (Jodrell/AO-search/AO-fold) & 1.58/1.95/1.67               & 1.67/1.13/-                      & 1.19/1.06/1.07                   & 2.34/1.01/0.98                   & 2.19/0.91/0.89\\
rms post-fit residuals ($\mu$s)                     & 4.87                              & 57.12                            & 13.92                            & 32.92                            & 13.31   
\enddata
\tablecomments{Numbers in parentheses are \tempo\ reported $1\sigma$ uncertainties.}
\tablenotetext{a}{95.4\% ($2\sigma$) upper limits.} 
\tablenotetext{b}{Assuming a pulsar mass of 1.4\,$M_\odot$ and an inclination of 90$\degrees$.}
\tablenotetext{c}{Assuming a pulsar mass of 1.4\,$M_\odot$ and an inclination of 60$\degrees$.}
\tablenotetext{d}{As implied by the \citet{cl02} NE2001 DM-distance model.}
\tablenotetext{e}{$L_{1400}=S_{1400}\,D^2$.}
%$^a$95.4\% ($2\sigma$) upper limits. \\
%$^b$Assuming a pulsar mass of 1.4\,$M_\odot$ and an inclination of 90$\degrees$. \\
%$^c$Assuming a pulsar mass of 1.4\,$M_\odot$ and an inclination of 60$\degrees$. \\
%$^d$As implied by the \citet{cl02} NE2001 DM-distance model. \\
%$^e$$L_{1400}=S_{1400}\,D^2$. \\
\end{deluxetable*}

\clearpage

\end{landscape}
\clearpage
%\end{turnpage}
%\global\pdfpageattr\expandafter{\the\pdfpageattr/Rotate 90}

\end{document}